\documentclass[preprint,12pt]{aastex}

\newcommand{\p}{\phantom{1}}
\usepackage{epsf}

\begin{document}

\title{Globular Cluster Systems in Four BCGs: \\ A262, A3560, A3565 and A3742
\footnote{Based on observations with the NASA/ESA Hubble Space Telescope,
obtained at the Space Telescope Science Institute, which is operated by the 
Association of Universities for Research in Astronomy, Inc., under NASA 
contract NAS 5-26555.  These observations are associated  with proposal ID
5910.}}

\author{Waldemar M. Oko\'n and William E. Harris}
\affil{Department of Physics and Astronomy, McMaster University, Hamilton,
ON L8S 4M1, Canada; okon@physics.mcmaster.ca, harris@physics.mcmaster.ca}

\begin{abstract}

We  have used deep  $I$-band (F814W)  images from  the HST  archive to
study  the  globular  cluster  systems around  the  brightest  cluster
galaxies (BCGs)  in Abell  262, 3560, 3565  and 3742.  Three  of these
BCGs  have inner  dust lanes  and peculiar  structural  features which
indicate past  histories of low-level interaction  and accretion.  The
deep $I-$band WFPC2 images have photometric limits which, for all four
galaxies, reach  near or just  beyond the GCLF turnover  point.  Their
specific frequencies are  $8.24 \pm 1.65$, $4.66 \pm  0.93$, $2.58 \pm
0.52$ and $2.62 \pm 0.52$ respectively,  all within a factor of two of
the normal  range for giant  ellipticals.  We obtain new  estimates of
the GCLF turnover magnitudes, which are shown to be consistent with an
adopted Hubble  constant of  $H_0 \simeq  70 \; {\rm  km \;  s^{-1} \;
Mpc^{-1}}$  on  the  ``Hubble  diagram''  of  GCLF  turnover  apparent
magnitude versus  redshift, on a distance scale  where the fundamental
GCLF calibrator  E galaxies  (M87 and others)  in Virgo are  at $d=16$
Mpc.

\end{abstract}



\section{Introduction}

Globular  clusters are  found in  their largest  numbers  within giant
elliptical  and cD  galaxies.  While  the clusters  constitute  only a
small  fraction of  a galaxy  halo's mass,  they are  the  most easily
identified  surviving  structures  of  the initial  stages  of  galaxy
formation.   This makes  them  important tracers  of galaxy  formation
models.   From  deep  wide-field  imaging,  several  features  of  the
globular cluster  system can readily be derived,  including the radial
distribution,  luminosity  distribution,   and  the  total  population
(specific frequency $S_N$).

In this paper  we obtain measurements of the  globular cluster systems
(GCSs)  in the  central  giant  E galaxies  within  four Abell  galaxy
clusters  in the redshift  range $cz  \sim 4000  - 4700$  km s$^{-1}$.
These are NGC 708 in A262, NGC 5193 in A3560, IC 4296 in A3565 and NGC
7014 in A3742.   The raw data are WFPC2  $I-$band exposures drawn from
the HST  archive.  Preliminary studies  of these systems were  made by
\citet{Lau98}, but  using only the PC1  chip data; we make  use of all
the WFPC2 data available, more than doubling the amount of information
on the globular cluster luminosity function (GCLF) in each galaxy, and
also enabling  us to study  the GCS radial distributions  and specific
frequencies  with much  higher  confidence.  Finally,  using the  GCLF
turnover magnitude  and redshift of  these galaxies and others  in the
literature, we  construct a purely GCLF-based Hubble  diagram and show
that it is consistent with $H_0 = 70$.


\section{Data Reduction}

The original  images were taken  in the F814W filter  by \citet{Lau98}
with  the WFPC2  camera, between  January and  April 1996  (Program ID
5910).  The target  galaxies were centered in the  high resolution PC1
chip, hence  maximizing the total globular  cluster population falling
within  the  field  of  view  of the  entire  camera.   The  exposures
(totaling  between 9200  and  16500 s)  were sub-pixel-shifted.   With
standard  IRAF  and  STSDAS   routines,  we  co-added  each  group  to
reconstruct  clean composite images  free of  bad pixel  artifacts and
cosmic ray contamination.

Various  small areas  on  each  image were  masked  out in  subsequent
analysis to eliminate contamination  from diffraction spikes of bright
stars,  or neighboring  galaxies in  the field.   Also,  the innermost
parts of the target galaxies  themselves were masked out, within radii
of  $3.''7$ , $3.''8$,  $5.''8$ and  $2.''8$ from  the centers  of the
galaxies respectively.

We  next subtracted background  light from  the combined  images.  For
this purpose  we used  a simple median  filter on the  WF2,3,4 fields,
where  the background  light gradient  is low.   However, for  the PC1
chip, we fitted elliptical isophotes (using STSDAS and ELLIPSE) to the
galaxy,  generating  synthetic  isophotal  contours  which  were  then
subtracted from the original.

With the galaxy  light subtracted, we found that  some of the galaxies
exhibited  various  types   of  interesting  substructure.   The  most
prominent of these  is the inner region of NGC  708, which exhibits an
easily  seen dust lane  (see the  illustrations of  each one  shown in
Lauer et al.).  Smaller dust  lanes were also detected within NGC 5193
and  IC  4296,  indicating  likely  accretion  events  in  the  recent
past. \citet{Gio82}  found that  A262, the host  cluster for  NGC 708,
contains  galaxies   deficient  of  HI   gas,  suggested  as   due  to
ram-pressure stripping.  This gas could  be accreted by the central cD
galaxy,  with its  large dark  halo.  A262  is a  spiral-rich cluster,
characterized by  a X-ray source, 3U 015+36,  which appears concentric
with the central  galaxy.  The gas and galaxy  virial temperatures for
the   cluster  have   recently  been   shown  to   be   comparable  by
\citet{Nei01}.  There seems to be  a general pattern for HI deficiency
to be associated with X-ray  emission, some other examples being A2147
in Hercules, or Coma.

Our procedures for  photometry of the four combined  images follow the
basic   sequence   outlined    in   more   detail   elsewhere   (e.g.,
\citet{WoHa00}, \citet{Kav00}).   Of the many  detected faint starlike
objects on the frames, the  vast majority are globular clusters within
the  BCGs.   At  these  distances  ($\sim$  55  Mpc)  they  appear  as
unresolved point sources,  even in the PC1 frames.   Hence, it is easy
to perform conventional point-spread  function (PSF) photometry on the
frames.   Independent  empirical PSFs  were  constructed from  several
moderately bright,  uncrowded stars in  each of the four  WFPC2 frames
with  DAOPHOT.  Then ALLSTAR  \citep{Ste94} was  used to  generate the
final photometry.  A standard detection threshold of 3.5 times the rms
scatter of  the sky  background was adopted  for the  image detection.
Crowding  at  all  magnitudes   was  completely  negligible  in  these
high-latitude fields.

An image classification  algorithm (CLASSIFY; defined by \citet{Kro80}
and \citet{Har_et91}) was then  used to calculate radial image moments
of the candidate objects and  thus to separate stellar from nonstellar
objects  in an objective  way.  By  using artificial-star  data passed
through   exactly  the  same   measurement  process,   we  established
boundaries for the CLASSIFY parameters  such that at least 95\% of the
true stars were retrieved at all magnitudes (for very similar examples
with illustrative  graphs, see \citet{WoHa00}  and \citet{Kav00}).  As
usual,  for  the  faintest  objects  the  image  moments  become  very
uncertain, and it becomes difficult to distinguish between stellar and
nonstellar   objects.    Also,  some   nonstellar   objects  will   be
accidentally classified  as stellar.  These  are statistically removed
from the  final GCLF by  subtracting a background  luminosity function
(see below).

Finally,   to  define   the  photometric   completeness   function  we
constructed  annular rings  around  the BCGs,  and employed  extensive
artificial-star tests to  measure the detection completeness $f(m,r)$,
as a function of magnitude  and radial distance.  In practice we found
that the  PC1 chip with its  strong background light  gradient was the
only area  where $f$ was a  significant function of radius;  on the WF
chips a single function $f(m)$ could be used.

The instrumental magnitudes were  converted to the Johnson-Cousins $I$
system  with  the  standard   transformations  for  $F814W$  found  in
\citet{Hol95}.   For  the four  individual  BCGs, Galactic  extinction
corrections of $A_I  =\;$ 0.16, 0.10, 0.11 and  0.06 respectively have
been  adopted ($A_I$  is given  by  $A_I=1.82 \;  E_{B-V}$, where  the
$E_{B-V}$  values   are  obtained  for  each   galaxy  from  NASA/IPAC
Extragalactic Database (NED)).  To employ the transformation equations
and also  to step back  and forth between  $I$ and the  (more normally
used)  $V$  magnitude scale  for  globular  clusters,  we have  simply
assumed  a  color  index  of  $(V-I)_0  = 1.1  \pm  0.1$,  typical  of
moderately  metal-rich globular  clusters  in giant  E galaxies  (e.g.
\citet{Kun99}).  The intrinsic range  in $(V-I)_0$, folded through the
transformation equations, will not introduce uncertainties larger than
$\pm 0.03$ in the calibration  of $I$.  The assumed $(V-I)_0$ value is
the mean  value representative of most  other gE galaxies,  and if the
GCSs were  entirely metal-rich or metal-poor, the  error introduced by
the assumption would be at most 0.1 mag.


\section{Analysis and Discussion}

\subsection{Radial Profiles}

\begin{figure}
\plotone{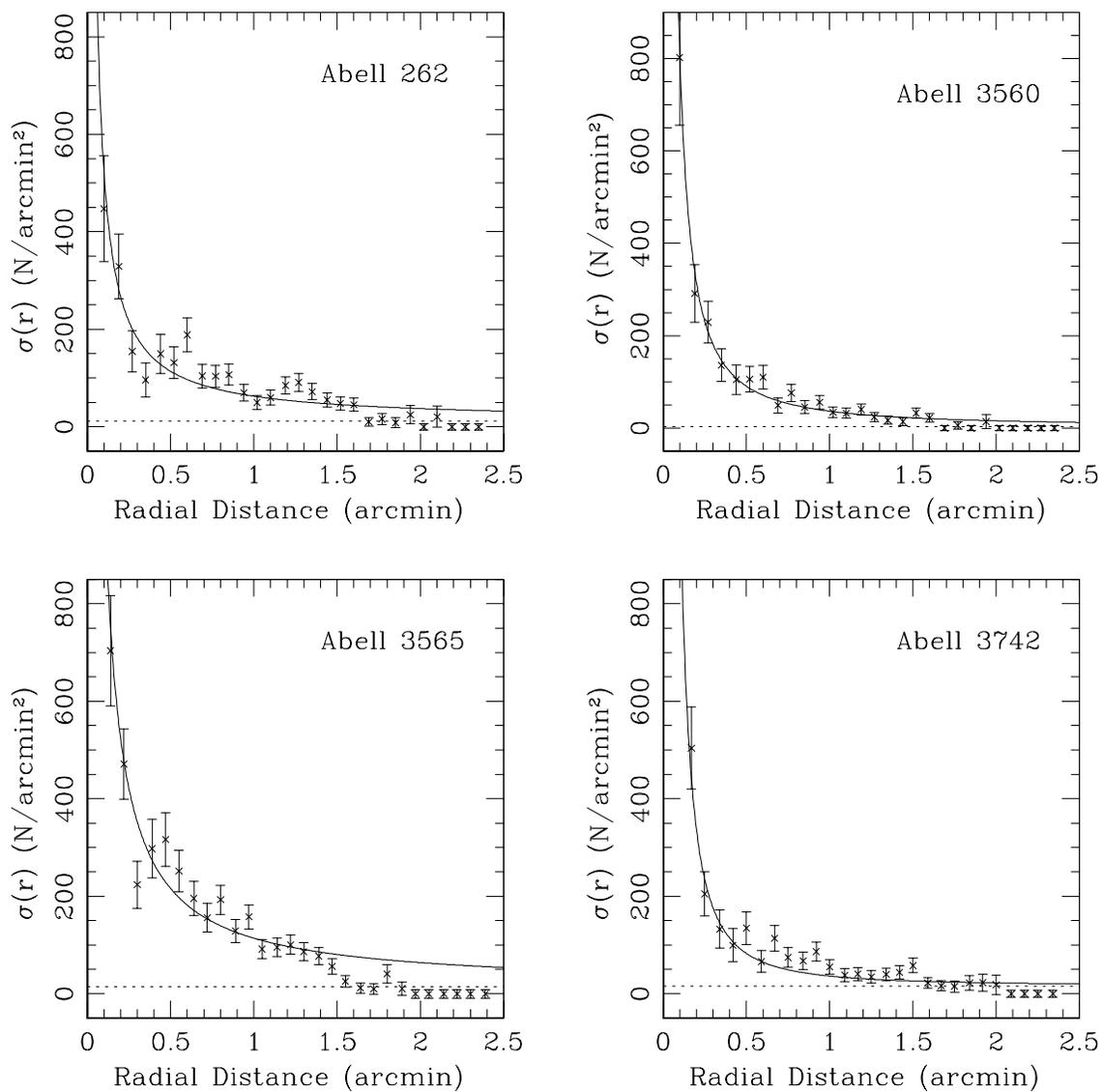}
\figcaption{Radial profiles of the detected starlike objects around
the four target BCGs.  The dotted lines indicate the adopted
background density levels, defined as the mean of the outermost
annuli.
\label{rad_plot}
}
\end{figure}

The projected number density  $\sigma$ of detected objects around each
galaxy plainly reveals an extensive GCS concentrated around the galaxy
center in each case.  The  profile is reasonably well represented by a
simple power-law form  $ \sigma(r) = \sigma_{cl}(r) +  \sigma_{bg} = a
\;r^b +  \sigma_{bg} $, where  $\sigma_{bg}$ is the  background number
density of starlike objects (mostly faint, small galaxies which passed
through  the  image classification  routines,  plus  a few  foreground
Galactic stars).  To obtain the profile parameters of the GCS for each
galaxy, we  subdivided the  WFPC2 fields into  annuli 50  pixels wide,
centered  on  the  BCGs.   The  number density  of  objects  was  then
calculated  down  to  a   cutoff  magnitude  at  which  incompleteness
corrections  were still small.  The projected  number density  is then
just  $\sigma =  N/A$, where  $N$ is  the number  of  detected objects
within each  annulus and $A$ is  the area of that  annulus which falls
within  the  WFPC2  boundaries  (minus the  small  masked-out  areas).
Completeness corrections, though small, were explicitly accounted for.
Finally,  the background  density $\sigma_{bg}$  on each  of  the four
fields was  defined as the mean  of the outermost  eight annuli, which
fall on the  outskirts of the WF chips.  This  corresponds to a radial
distance greater  than $105''$ (or about  30 kpc) from  the centers of
the BCGs.  Although the GCSs  probably extend at trace amounts farther
out  than  this boundary,  the  directly  observed $\sigma(r)$  curves
(Figure~\ref{rad_plot})  have   plainly  almost  leveled   off  there,
indicating that we are already including the main portion of the GCS.

\begin{figure}
\plotone{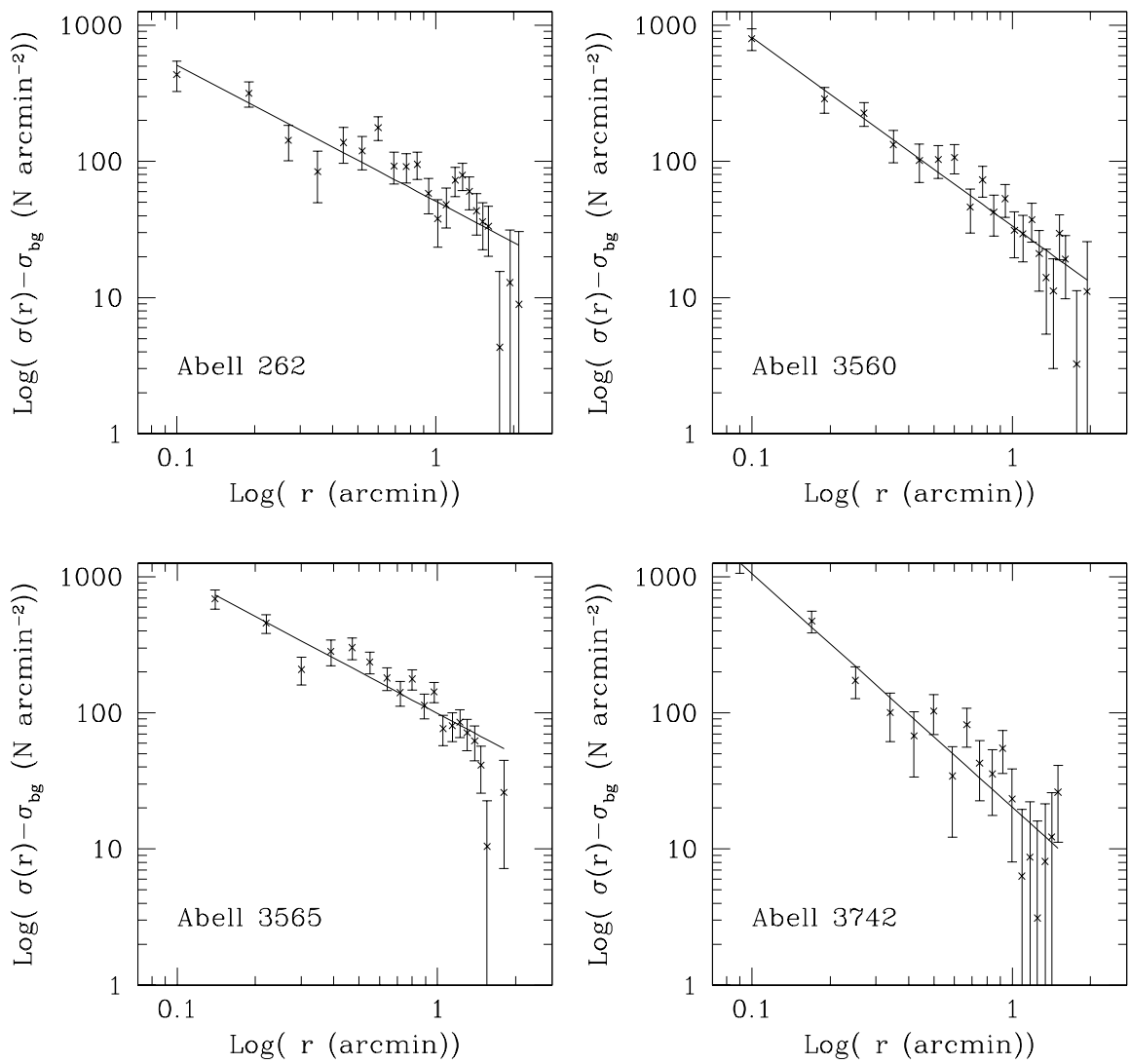}
\figcaption{Log-log  plot of  radial  profiles of  the detected  starlike
objects  around  the  four   target  BCGs.   The  background  density,
$\sigma_{bg}$, has been subtracted.
\label{lograd_plot}
}
\end{figure}

Figure~\ref{lograd_plot}     shows     log-log     plots    of     the
background-subtracted  density   profiles  for  the   four  BCGs,  and
Table~\ref{rad_table} lists our deduced values for the power-law index
$b$, as well  as the adopted $\sigma_{bg}$.  We  note the bumpiness of
the $\sigma_{cl}(r)$  curve for NGC 708  in Abell 262.   It exhibits a
clear peak  around $1.'25$, and a  less clear peak  around $0.'6$.  If
not  simply statistical  fluctuations, these  could be  the  result of
uncompensated background contamination (clumpiness in the distribution
of  faint,  ultra-distant  galaxy  clusters,  or  even  some  residual
globular cluster  populations which belong to  neighboring galaxies in
A262;  there are  several smaller  elliptical  galaxies as  well as  a
companion spiral in the WFPC2 field).

Also, the  slope of  the $\sigma_{cl}(r)$ curve  for IC 4296  in A3565
declines smoothly out to about  $1.'5$, and then exhibits a relatively
sharp  drop-off.  This  effect  is  not apparent  in  the other  three
systems.  It  is possible that this  feature, if real (and  not just a
prosaic  result of  overestimating the  background count),  is  due to
tidal truncation.  There are no large nearby galaxies in the immediate
vicinity of IC 4296. However,  \citet{Mul96} in a {\sl ROSAT} study of
groups and poor clusters find that the X-ray contours in A3565 are not
centered on the BCG, but rather  at a point roughly halfway between IC
4296 and the neighboring spiral  IC 4299.  The mass that \citet{Mul96}
have derived for  the X-ray gas is about one-third of  the mass of the
entire   cluster  derived  from   the  galaxy   velocity  distribution
\citep{Wil99}.  It might therefore be possible that the BCG oscillates
about the center of the potential  well of the cluster, which could be
responsible for the truncation of its halo.

For luminous  giant E  galaxies like these,  the GCS profile  shape is
expected to be rather flat, becoming more shallow as galaxy luminosity
increases.    A  general   relation   for  the   power-law  slope   is
\citep{Kai96}.
$$ b \, = \,  -0.29 \, M_V^T \, - \, 8.00 $$ The  only one of our four
target  BCGs  which  deviates  fairly  strongly  from  this  empirical
relation is NGC  708 (an observed slope $b =  -1.0$ versus an expected
one of $-1.7$).  It is, however, the  only one in our sample with a cD
envelope,  which  has clearly  made  it  more  radially extended  than
normal.

\begin{table}[htbp]
\caption{Radial Density Profile Data} 
\begin{center}
\begin{tabular}{lccc}
\hline\hline

Galaxy             &        $b$        &    $\sigma_{bg}$   \\ \hline
N708  \p(A262)     & -1.00 $\pm$ 0.11  &   11.6 $\pm$ 3.1   \\ 
N5193 (A3560)      & -1.38 $\pm$ 0.07  & \p 2.8 $\pm$ 1.9   \\
I4296 \p(A3565)    & -1.02 $\pm$ 0.08  &   15.0 $\pm$ 3.9   \\
N7014 (A3742)      & -1.71 $\pm$ 0.11  &   16.0 $\pm$ 3.6   \\ \hline

\label{rad_table}
\end{tabular}
\end{center}
\end{table}


\subsection{Globular Cluster Luminosity Function}

The  globular  cluster luminosity  function  (GCLF) is  conventionally
defined as  the number  of globular clusters  per unit  magnitude.  In
this  form it  has  a unimodal  and  nearly symmetric  shape, and  for
analytical  purposes  can be  reasonably  approximated  by a  Gaussian
function (cf.  \citet{Har01} for extensive discussion and background).
For giant  elliptical galaxies  the Gaussian dispersion  $\sigma_V$ of
the distribution is found to be  $1.36 \pm 0.03$ (see below) while the
peak frequency,  also known  as the GCLF  turnover point, is  equal to
$M_V^0 =  -7.33 \pm  0.04$ on a  distance scale where  the calibrating
ellipticals (M87 and others in the  Virgo cluster) are at $d = 16$ Mpc
(see   \citet{Har01};    \citet{Kav00}).    The   demonstrated   small
galaxy-to-galaxy spread  in $M_V^0$  makes it an  interesting standard
candle  for  distance  measurements,   because  it  is  detectable  to
distances beyond  which peculiar motions  of galaxies bias  the Hubble
flow (see \S 3.5).

To   plot   the   GCLFs   for    our   four   GCSs,   we   have   used
completeness-corrected  totals  in  0.25   mag  bins  above  the  50\%
completeness  level magnitudes.  A  background GCLF  was statistically
subtracted, defined  from the objects (mostly  background galaxies) in
the outskirts  of the fields  (same region as  the one used  to define
$\sigma_{bg}$  above).  The  best-fit Gaussian  function  was obtained
with a constrained $\chi^2$ minimization method.

When  solving for  the  two important  GCLF  parameters (the  apparent
magnitude  of the  turnover, in  this case  $I^0$, and  the dispersion
$\sigma_I$),  it  is  important  to  note  that  their  solutions  are
correlated  (see \citet{SeHa93}; \citet{HaWh87};  \citet{Kav00}).  If,
as is true in our case,  the photometric limit $I(lim)$ is close to or
just  past the  true turnover  magnitude,  the fitted  values for  the
turnover and dispersion are constrained from only one side (the bright
end).   In this situation,  attempts to  solve simultaneously  for the
turnover  and the  dispersion tend  to produce  overestimates  of both
quantities.  One  can obtain a {\it systematically}  more accurate fit
by adopting a fiducial value  for the dispersion, and solving only for
the  turnover   magnitude.   Fortunately,  it  has   been  shown  (see
\citet{Whit96} and  \citet{Har01}) for more  than a dozen  gE galaxies
with well measured GCLFs, that  the dispersion is very consistent from
one galaxy to another, with  $\langle \sigma_V\rangle = 1.36 \pm 0.10$
rms scatter.  We adopt that dispersion value here and solve only for
the turnover magnitudes $I^0$.

\begin{figure}
\plotone{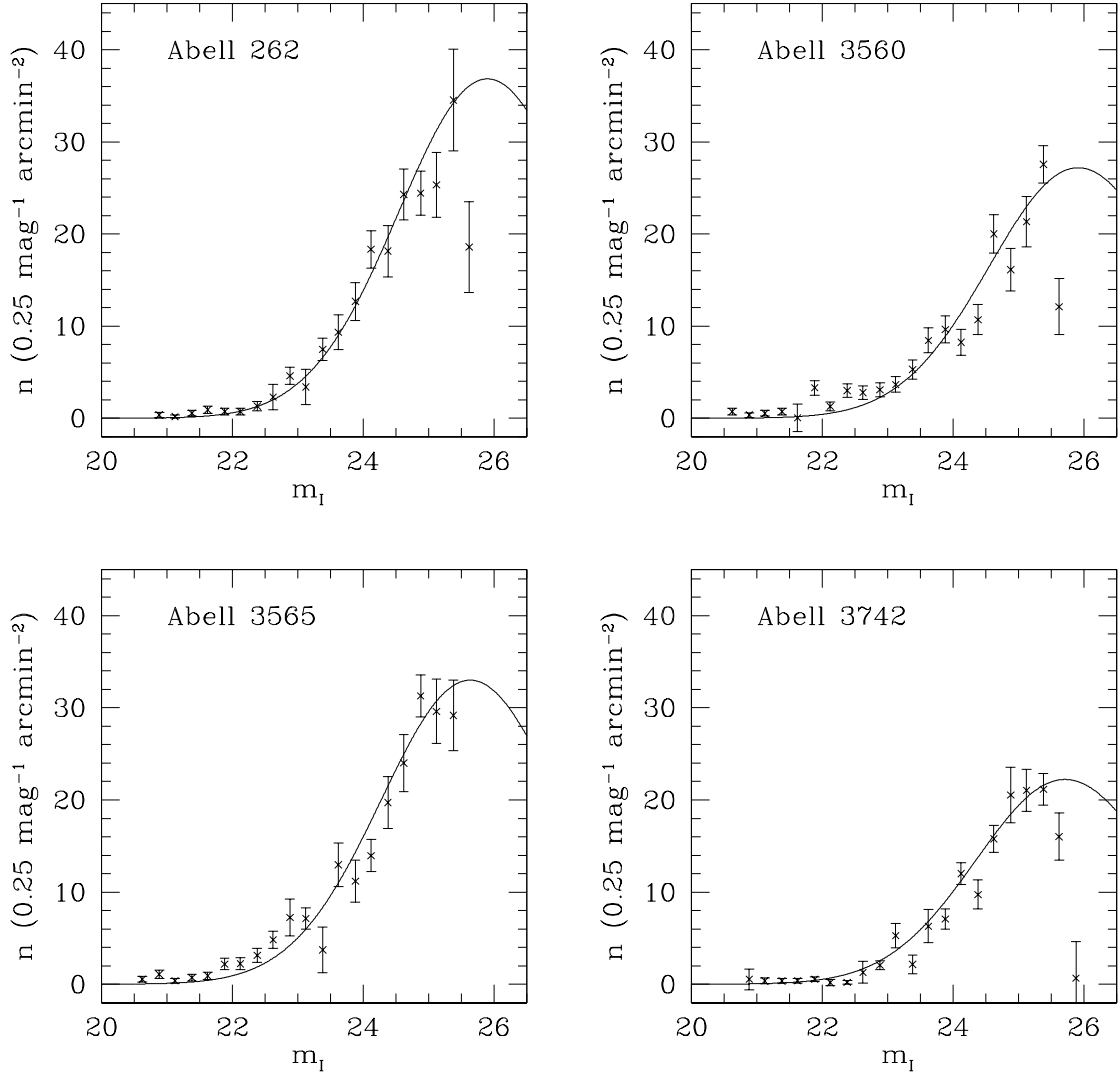}
\figcaption{The data defining the globular cluster luminosity function
(GCLF) for bins brighter than the 50\% completeness level.  The solid lines are
Gaussians of width $\sigma_V=1.36$.
\label{gclf_plot}
}
\end{figure}

Our  full completeness-corrected and  background-subtracted luminosity
functions,  along   with  the  fitted   Gaussian  functions  (assuming
$\sigma_I   =  1.36$)  are   shown  in   Figure~\ref{gclf_plot}.   The
individual points  were weighted as $(n/e(n))^2$ where  $n$ and $e(n)$
denote the number of objects  in the bin and the internal uncertainty.
Table~\ref{gclf_table}  lists  the  resulting turnover  $I$-magnitudes
obtained for the GCLFs, with successive columns giving the result from
(a) only  the PC1 data, (b) only  the WF data, (c)  all data combined,
and finally (d) the values  determined previously by Lauer et al.  The
$V^0$ values result from the fitted $I^0$ once we add our adopted mean
color index $(V-I)_0  = 1.1 \pm 0.1$.  In the  Table, the quoted $I^0$
values   have  had   Galactic  extinction   subtracted.    The  quoted
uncertainties  in   the  Table  represent  only   the  {\sl  internal}
uncertainty of the fitting procedure.

Our results for the four  BCGs yield turnover magnitudes quite similar
to those obtained  by \citet{Lau98} for A262, A3742,  but fainter than
their values for A3560 and A3565. Lauer et al. employed the same basic
fitting curve  (a Gaussian with dispersion 1.4  magnitudes), but there
are two notable differences between their analysis and ours. First, we
have used all  the available WFPC2 data (not just  the PC1 data), more
than doubling the total cluster  populations used in the fit.  Second,
there are  important differences of  detail in the  fitting procedure.
We define an {\sl empirical} background LF from the outer parts of the
images; then  subtract this background  from the total  number counts;
and then fit  the (residual) LF with the  model Gaussian function.  By
contrast, Lauer et al. fit a model to the raw data which includes both
the assumed Gaussian GCLF shape  and a model background LF which rises
smoothly and  exponentially with magnitude.   Either procedure depends
for  its accuracy  on the  correctness  of the  assumed background  LF
particularly at faint levels.   Thus, for example, if their background
model overestimated  the true  background at or  beyond the  true GCLF
turnover, it would  yield an $I^0$ which was  artificially too bright.
In cases  such as these where  the photometric limit is  very close to
the  actual turnover,  the  true (external)  uncertainty  in $I^0$  is
likely to be closer to  $\pm 0.3$ magnitude (cf.\ the references cited
above).   However, primarily because  of the  much larger  GCLF sample
size in  our data and  our locally determined backgrounds,  we believe
our  turnover  determinations to  be  improvements  over the  previous
work.\footnote{We  have compared  the  GCLFs from  only  our PC1  data
directly  with those of  Lauer et  al.  (see  their Figure  8).  After
correction for the different bin sizes, we find that their actual GCLF
datapoints and ours  agree quite closely, to well  within the internal
errors of either  set.  Our PC1 data do not reach  as faint as theirs,
however,   indicating   that   our   detection  threshold   was   more
conservative.}

\begin{figure}
\plotone{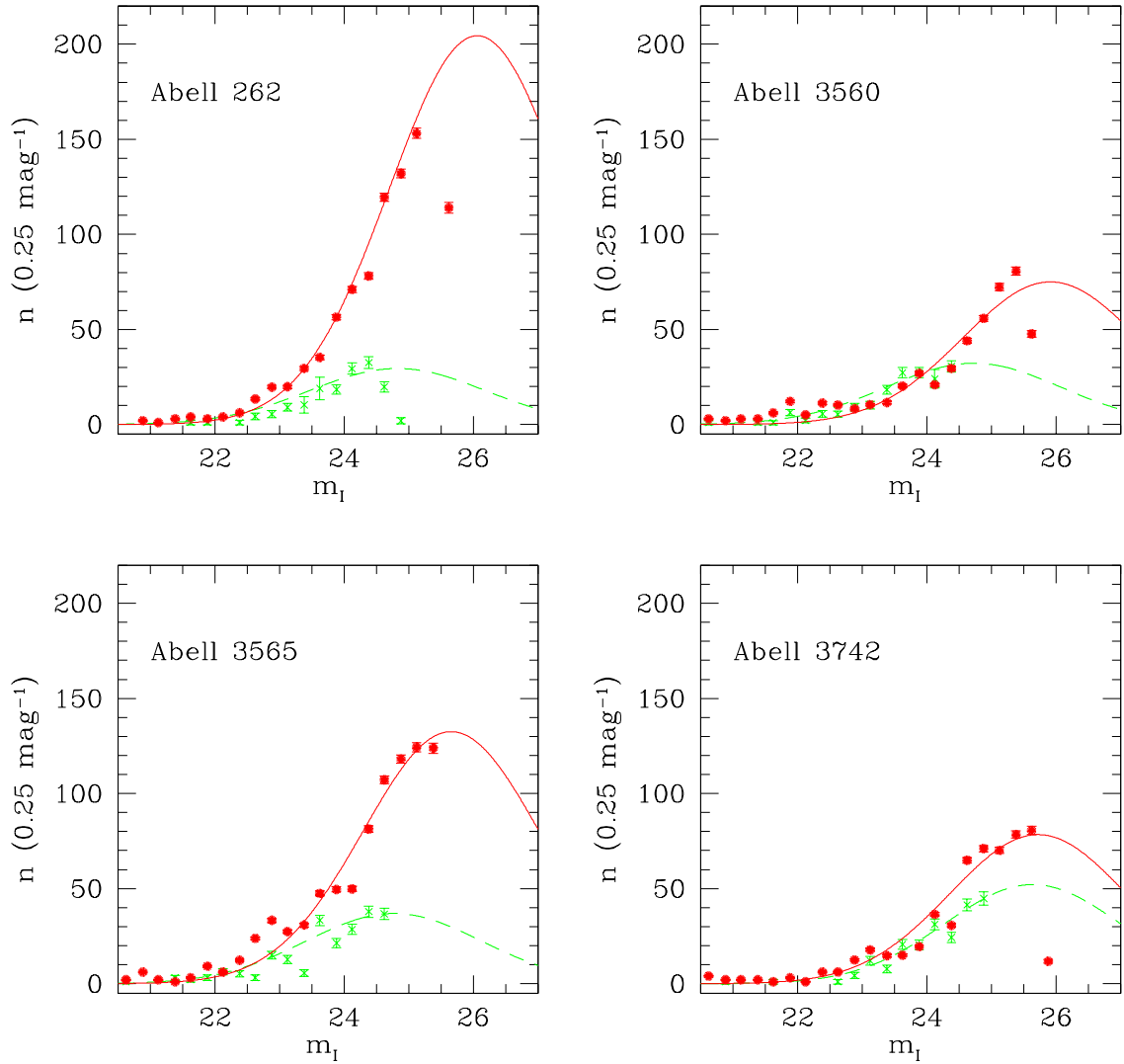}  
\figcaption{The   data  defining   the  globular   cluster  luminosity
functions (GCLFs) for bins above the 50\% completeness level on the PC1
(dashed) and WF (solid) chips.  The fitted lines are Gaussians of width
$\sigma_V=1.36$.
\label{gclf_comp}
}
\end{figure}

In  Figure~\ref{gclf_comp}  we show  the  GCLF  from  our data  broken
separately  into  the  WF  and  PC1 data  (and  with  separate  fitted
Gaussians  for both  subsets).  Notably,  the fits  from the  PC1 data
alone  yield consistently  brighter turnovers  than do  the  deeper WF
data, indicating the importance  of obtaining photometric limits at or
beyond the  actual turnover to  avoid systematic errors.  The  WF data
have limiting  magnitudes similar  to those of  Lauer et al.,  and the
turnover magnitudes  we deduced from  the WF data alone  are generally
closer to those  deduced by Lauer et al., as are  those from the total
(PC+WF) dataset.  These latter fits are certainly to be preferred over
the PC1 data alone.

We note  that the datapoints in  the faintest bin for  A262, A3560 and
A3742 in  Figure~\ref{gclf_plot} and in  the WF Figure~\ref{gclf_comp}
fall clearly below the fitted GCLF  curve.  This does not appear to be
due to an  incorrect completeness correction at that  level.  The only
other obvious  possibility is that  our empirical background  has been
overestimated in the  last bin, but it is unclear  whether that is the
case without deeper data to draw on.

\begin{table}[htbp]
\caption{GCLF Turnover Magnitudes}
\begin{center}
\begin{tabular}{lcccc}
\hline\hline

Galaxy           & $I^0$(PC1)&  $I^0$(WF) & $I^0$(total)        & $I^0$(Lauer et al.)      \\ \hline
N708  \p(A262)   &   24.66   &    25.90   &  25.74  $\pm$  0.17 &     25.85 $\pm$ 0.25     \\
N5193 (A3560)    &   24.62   &    25.81   &  25.81  $\pm$  0.18 &     25.12 $\pm$ 0.25     \\
I4296  \p(A3565) &   24.66   &    25.72   &  25.54  $\pm$  0.16 &   \p24.72 $\pm$ $0.25^1$ \\
N7014 (A3742)    &   25.58   &    25.71   &  25.66  $\pm$  0.19 &     25.77 $\pm$ 0.25     \\ \hline

\multicolumn{5}{p{12cm}}{$^1$ The turnover magnitude for IC 4296 listed by
 \citet{Lau98} in their Table 2 ($I_0=25.72$) is a typographical error.  It
 should read $I^0=24.72$, which is the value from their graph (Figure 8)
 \citep{Lau01}.}
\label{gclf_table}
\end{tabular}
\end{center}
\end{table}


\subsection{Luminosity and Mass Distribution Functions}
 
The luminosity distribution function  (LDF), or number of clusters per
unit  {\it luminosity}  ($dN/dL$),  is the  visible  signature of  the
cluster  mass   distribution  and   is  a  more   physically  oriented
representation of  the GCLF.   In this plane  it usually appears  as a
rough  power   law,  $dN/dL   \sim  L^{-\alpha_L}$.   By   adopting  a
mass-to-light  ratio   one  can  also  immediately   obtain  the  mass
distribution  function  $dN/dM  \sim M^{-\alpha_M}$,  where  $\alpha_L
\simeq \alpha_M$ as long as  the $M/L$ ratio is roughly independent of
cluster mass (see \citet{Man91} and \citet{McL00}).

It has  been noted many  times that within  observational uncertainty,
the  globular cluster  mass distribution  follows a  simple  power law
relation which  has the same shape  as the mass  distribution of giant
molecular  clouds (GMCs)  in large  spirals, cloud  cores  embedded in
GMCs, and giant HII regions  in large spirals (e.g. \citet{HaPu94} and
subsequent papers).  The power-law slope $\alpha_M$ falls consistently
in the range $1.8 \pm 0.2$ for $L > 10^5 L_{\odot}$, a value which can
be  at least  approximately  explained by  cluster formation  pictures
ranging   from   collisional  growth   of   protocluster  gas   clouds
\citep{McLPu96} to turbulence spectra \citep{Elm97}.

Here we define  the LDFs observationally in a  separate procedure from
the GCLF: we  convert the apparent magnitudes of  the detected objects
to  absolute magnitudes  $M_V$ using  the galaxy  distance  moduli and
adopted cluster color, and then to luminosities $L_V$.  The background
LDF is  subtracted and the data  binned in steps  of $10^5 L_{\odot}$.
For the  mass distribution function we have  multiplied the luminosity
values  by a  mass-to-light  ratio $M/L_V  =  1.45$ \citep{McL00}  and
binned the data in steps of $10^5 M_{\odot}$.

We note that the lowest luminosity  (and hence mass) point for each of
the functions has been excluded  from the fits.  This corresponds to a
point $L < 10^5 L_{\odot}$, since at this luminosity, the slope of the
power  law  relation  changes   for  the  distribution  function  (see
\citet{McL94}).

\begin{figure}
\plotone{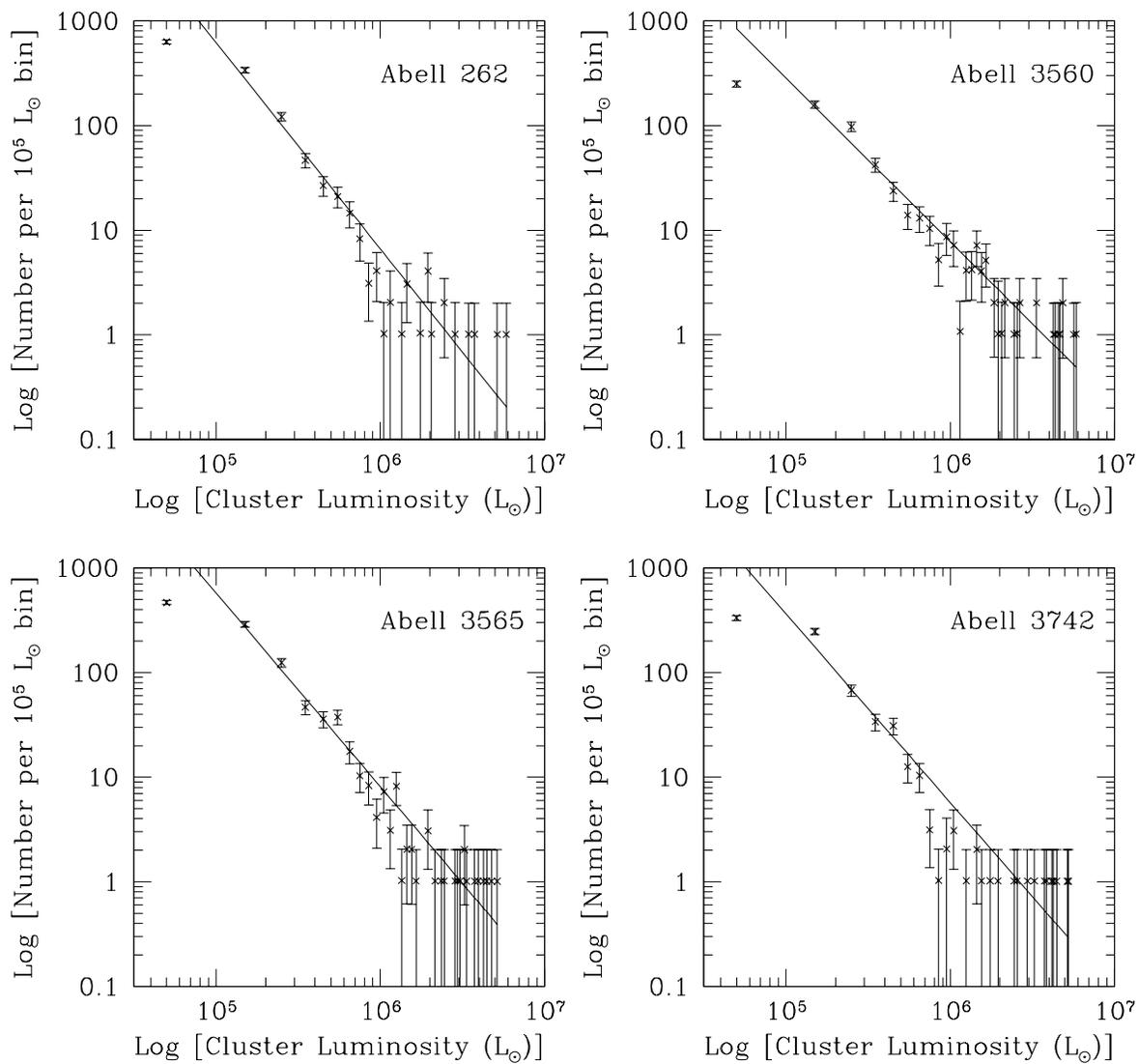}
\figcaption{The  data defining  the  luminosity distribution  function
(LDF).  The solid lines are  the least squares fits, with slopes given
in  Table~\ref{ldf_table}.   The  lowest  luminosity  point  has  been
excluded from the fit in all cases (see text).
\label{ldf_plot}
}
\end{figure}

Figure~\ref{ldf_plot} shows  the LDFs which  we have obtained  for the
four  galaxies.  Table~\ref{ldf_table} lists  the values  obtained for
the  slopes of  the power  law relations  of the  luminosity  and mass
distribution  functions,   $\alpha_L$  and  $\alpha_M$.    (These  are
slightly but not  significantly different in each pair  because of the
different bin boundary locations in  $L$ and $M$.)  With the exception
of NGC  5193 in A3560  which has a  value of $\alpha_M$  notably lower
than the expected $1.8 \pm 0.2$, the other three galaxies are entirely
consistent  with the  other  giant ellipticals  in  Virgo, Fornax  and
elsewhere.

\begin{table}[htbp]
\caption{Luminosity and Mass Distribution Function Data}
\begin{center}
\begin{tabular}{lccc}
\hline\hline

Galaxy           &     $\alpha_L$     &          $\alpha_M$     \\ \hline
N708  \p(A262)   &  -1.97 $\pm$ 0.11  &       -1.86 $\pm$ 0.08  \\
N5193 (A3560)    &  -1.57 $\pm$ 0.07  &       -1.48 $\pm$ 0.05  \\
I4296  \p(A3565) &  -1.85 $\pm$ 0.07  &       -1.74 $\pm$ 0.07  \\
N7014 (A3742)    &  -1.96 $\pm$ 0.10  &       -1.91 $\pm$ 0.08  \\ \hline

\label{ldf_table}
\end{tabular}
\end{center}
\end{table}


\subsection{Total Populations and Specific Frequencies}

Next we estimate the total populations and specific frequencies of the
GCSs.  Here $S_N$ is the number of clusters per unit galaxy luminosity
\citep{Har_vdB81, Har91}:
$$S_N = N_{cl}  \times 10^{0.4 \; (M_V^T+15)}$$ where  $N_{cl}$ is the
total  number of  clusters,  and $M_V^T$  is  the integrated  absolute
magnitude  of the  host galaxy.   $S_N$ is  found to  differ  among gE
galaxies by more than an  order of magnitude \citep{HHM98, Har01}.  It
is  known that  for cD-type  and BCG  galaxies the  specific frequency
increases systematically with the total galaxy luminosity, the size of
the  surrounding cluster  of galaxies,  and  the X-ray  halo gas  mass
\citep{BTM97, Bla99, HHM98, McL99, Kav99}, albeit with a factor-of-two
scatter that is not understood in detail.

We calculate  the total number  of clusters by integrating  the radial
profiles and scaling  up the result by the fraction  of the area under
the bright half of the  GCLFs (the completeness magnitude used for the
radial profiles corresponds to  90\% completeness, while for the GCLFs
it   is   50\%).     Following   previously   established   convention
\citep{Har01}, the  result is  then doubled, which  implicitly assumes
that  the GCLF  is symmetric  about  the turnover  magnitude.  We  can
therefore think of  specific frequency as equivalent to  the number of
{\it bright} clusters in a galaxy.  Also, the value of $S_N$ is fairly
insensitive to the assumptions in the galaxy distance, because changes
in  distance will affect  the calculated  galaxy luminosity  and total
cluster population in the same sense (see \citet{Har_vdB81}).

\begin{figure}
\plotone{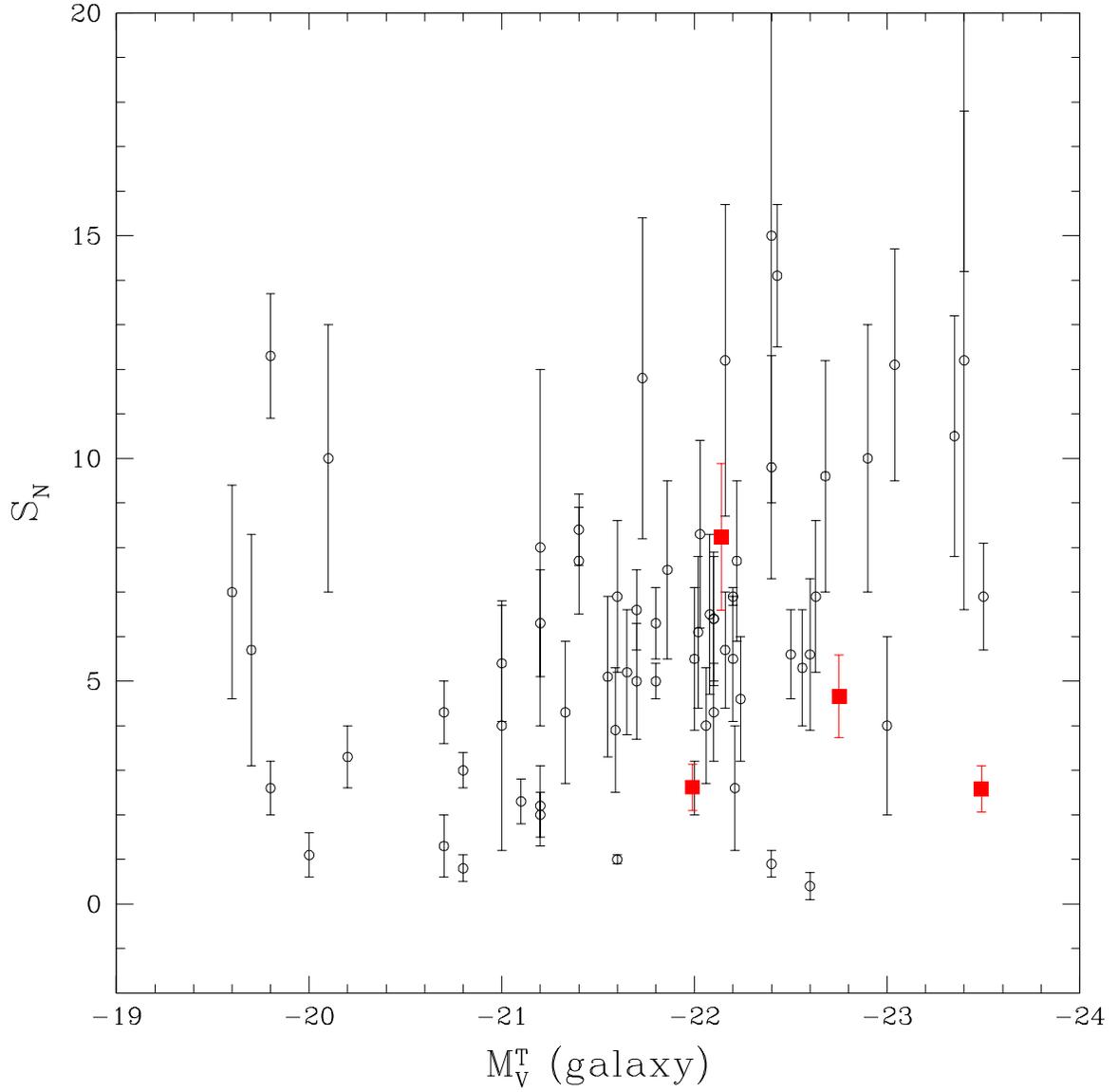}  
\figcaption{Specific  Frequency $S_N$  plotted against  luminosity for
elliptical  galaxies.  Squares  represent  our results  for the  BCGs,
while open circles are results for other BCGs and gEs.
\label{sn_plot}
}
\end{figure}

The total  populations along with  the specific frequencies  $S_N$ are
listed   in  Table~\ref{sn_table}.   Figure~\ref{sn_plot}   shows  our
results along with those from other gE galaxies previously published.

NGC 708 in  A262 has the highest value for  $S_N$, consistent with the
fact that it  is the only genuine cD-type galaxy  of the four studied.
The  other three  galaxies are  central BCGs  without the  extended cD
envelope, and have lower values  for $S_N$.  Somewhat of an outlier is
IC 4296 in A3565, with a $S_N$ value of 2.6, definitely on the low end
for  a such  a  high-luminosity elliptical.   Its  $S_N$ is,  however,
similar to  those of ``field'' ellipticals which  are commonly thought
to have  formed by major mergers between  relatively less cluster-rich
disk galaxies.   Even if  the merging galaxies  are quite  gas-rich, a
high$-S_N$ elliptical  would not  necessarily result, since  new field
stars {\sl and} star clusters both form during the merger, and the net
ratio of  clusters to  field stars in  the final merger  product could
either increase or decrease.  The  final $S_N$ would be higher only if
the efficiency of cluster formation was considerably enhanced over the
cluster formation that took place in the original protogalactic epoch.
In actually  observed cases of recent disk/disk  mergers, what appears
to be emerging  in every case is an elliptical with  $S_N \sim 2$ (see
\citet{Har01} for more extensive discussion).

\begin{table}[htbp]
\caption{Specific Frequency Data}
\begin{center}
\begin{tabular}{lccc}
\hline\hline 

Galaxy           &  $M_V$  &     $N_{cl}$      &         $S_n$     \\ \hline
N708  \p(A262)   & -22.14  &  5924 $\pm$  302  & 8.24  $\pm$ 1.65  \\       
N5193 (A3560)    & -22.75  &  5878 $\pm$  355  & 4.66  $\pm$ 0.93  \\       
I4296  \p(A3565) & -23.49  &  6400 $\pm$  265  & 2.58  $\pm$ 0.52  \\       
N7014 (A3742)    & -21.99  &  1634 $\pm$ \p 88 & 2.62  $\pm$ 0.52  \\ \hline

\label{sn_table}
\end{tabular}
\end{center}
\end{table}

\subsection{The Hubble Diagram}

By  using our  results for  the GCLF  turnover magnitudes,  along with
previous results for  several other gE galaxies from  the literature a
classic ``Hubble  Diagram'' of redshift  $cz$  against  the  
apparent  turnover magnitude $V^0$ of the GCLF can be 
constructed  (see \citet{Har01} and
\citet{Kav00} for the first use  of this diagram for GCLFs).

Hubble's  law  states $cz=H_0  d$.   This  can  be rewritten as:

$$ {\rm log}(cz)  = 0.2 V^0  + {\rm log}H_0 -0.2M_V^0  -5 $$

where $M_V^0$ is  the GCLF turnover luminosity {\it  for gE galaxies},
and $H_0$  is expressed in  the usual units  of ${\rm km \;  s^{-1} \;
Mpc^{-1}}$.  Plotting  log($cz$) against the  apparent magnitude $V^0$
gives a  straight line  of slope 0.2,  and a zeropoint  which contains
$H_0$ and $M_V^0$.

\begin{figure}
\plotone{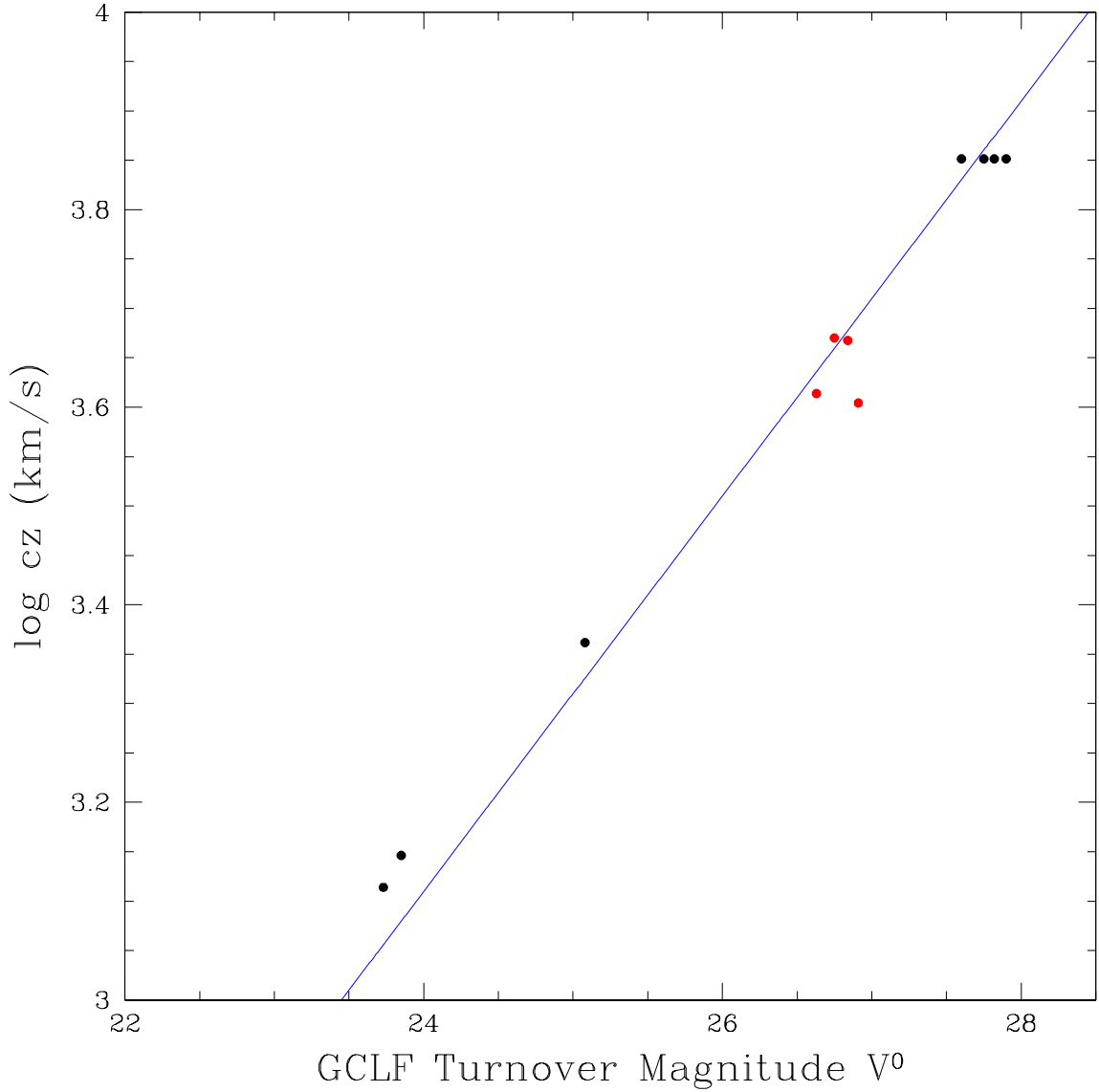}
\figcaption{Hubble diagram for GCLF turnover magnitudes.
The galaxy redshift $cz$
is plotted against the apparent magnitude of the GCLF turnover, $V^0$.
The solid line is the expected relation for $H_0 = 70$ and 
an absolute turnover magnitude $M_V^0 = -7.33$ (see text).
\label{hub_plot}
}
\end{figure}

The available  data for a total  of eleven BCG galaxies  or groups are
listed in  Table~\ref{TO_table} and plotted  in Figure~\ref{hub_plot}.
The values  for the  Virgo and Fornax  clusters are the  weighted mean
$\left<V^0\right>$   values   of    the   individual   galaxies   (see
\citet{Kav00}).  The  mean radial velocities  of each galaxy  or group
are   taken    from   \citet{Fab89},   \citet{Gir93},   \citet{Huc88},
\citet{BPT93},   \citet{Ham96},  \citet{CoDu96},   and  \citet{Lau98}.
Also, the recession  velocities $cz$ for the target  galaxies assume a
Local Group  infall to Virgo  of $250 \pm  100 \; {\rm km  \; s^{-1}}$
(e.g.,  \citet{Ford96};   \citet{Ham96};  \citet{JeTa93},  among  many
others).

The  four galaxies  we have  investigated  here fall  well within  the
pattern established by  the others (from Virgo at  low redshift out to
Coma at the  highest redshift).  A value of $H_0$  near $\simeq 70$ km
s$^{-1}$   Mpc$^{-1}$,  which   represents  a   consensus   of  recent
determinations \citep{Fre01}, along with $M_V^0 = -7.33$ \citep{Kav00,
Har01},  matches the total  range of  points within  their measurement
uncertainties.

\begin{table}[htbp]
\caption{GCLF turnover magnitudes in distant ellipticals}
\begin{center}
\begin{tabular}{lcccc}
\hline\hline 

Cluster & Galaxy & Redshift $cz$ (km/s) &  $V^0$ (GCLF) &   Source \\ \hline
Virgo   &  6 gEs &    1300              &  $23.71 \pm 0.03$ & 1    \\
Fornax  &  6 gEs &    1400              &  $23.85 \pm 0.04$ & 1    \\
NGC 5846&NGC 5846&    2300              &  $25.08 \pm 0.10$ & 2    \\
Coma    & IC 4051&    7100              &  $27.75 \pm 0.20$ & 3    \\
Coma    &NGC 4874&    7100              &  $27.82 \pm 0.12$ & 1    \\
Coma    &NGC 4881&    7100              &  $> 27.6$         & 4    \\
Coma    &NGC 4926&    7100              &  $27.90 \pm 0.20$ & 5    \\
A 262   &NGC 708 &    4650              &  $26.84 \pm 0.17$ & 6    \\ 
A 3560  &NGC 5193&    4020              &  $26.91 \pm 0.18$ & 6    \\
A 3565  & IC 4296&    4110              &  $26.63 \pm 0.16$ & 6    \\
A 3742  &NGC 7014&    4680              &  $26.75 \pm 0.19$ & 6    \\ \hline

\label{TO_table}
\end{tabular}
\end{center}
Sources: (1) \citet{Kav00}; (2) \citet{For96}; (3) \citet{WoHa00};
         (4) \citet{Bau95};  (5) \citet{JJp}; (6) this paper
\end{table}

\section{Summary}

Using deep  $I$-band photometry from  the $HST$/WFPC2 archive  we have
studied the globular cluster systems in NGC 708, NGC 5193, IC 4296 and
NGC 7014,  the BCGs in A262,  A3560, A3565 and  A3742.  The photometry
allowed us to construct  the globular cluster luminosity functions for
each  of the  above  cluster systems,  reaching  to or  just past  the
turnover point.

We have used a constrained  $\chi^2$ method to fit a Gaussian function
to  each of  the four  GCLFs, adopting  a width  $\sigma_V=1.36$.  The
resulting  background  and   extinction  corrected  $V$-band  turnover
magnitudes were found  to be at $V^0$ = 26.84  $\pm$ 0.17, 26.91 $\pm$
0.18, 26.63  $\pm$ 0.16, 26.75 $\pm$ 0.19  respectively.  These values
improve  on previous  results  of \citet{Lau98}.   The Hubble  diagram
generated  from our GCLF  turnover data  combined with  other material
from the  literature matches $H_0  \simeq 70 \;  {\rm km \;  s^{-1} \;
Mpc^{-1}}$.

We have  obtained the luminosity  and mass distribution  functions for
the GCSs.   The slopes  of the  power law relations  were found  to be
consistent with  those for other  giant elliptical galaxies,  with the
exception of NGC 5193 in A3560.

The total  cluster populations  $N_{cl}$ and the  specific frequencies
$S_N$ for each  system were calculated.  The specific  frequency of IC
4296 in A2565  is surprisingly low for its  high luminosity, while the
values for the other three  galaxies fall within the established trend
of specific frequency versus galaxy luminosity.

\acknowledgements

We would like to thank  J.J.Kavelaars and Marcel VanDalfsen for useful
comments and advice.  This work  was supported by the Natural Sciences
and Engineering Research Council of Canada through a research grant to
WEH.

\clearpage



\end{document}